\documentstyle[12pt,epsfig]{article}
\newcommand{\be}{\begin{equation}}
\newcommand{\ee}{\end{equation}}

\def\simlt{\lower.5ex\hbox{\ltsima}}
\def\gtsima{$\; \buildrel > \over \sim \;$}
\def\simgt{\lower.5ex\hbox{\gtsima}}

\def\simlt{\lower.5ex\hbox{\ltsima}}
\def\gtsima{$\; \buildrel > \over \sim \;$}
\def\simgt{\lower.5ex\hbox{\gtsima}}

\def\cm{{\rm\,cm}}

\def\ergs{\ {\rm erg~s^{-1} }}
\def\ergcm2{\ {\rm erg~cm^{-2} }}
\def\ergscm2{\ {\rm erg~s^{-1}~cm^{-2} }}

\def\cm2s{\ cm^2 ~s^{-1} }

\def\s{\ifmmode \widetilde \else \~\fi}
\def\={\overline}

\def\spose#1{\hbox to 0pt{#1\hss}}

\def\lta{\mathrel{\spose{\lower 3pt\hbox{$\mathchar"218$}}
     \raise 2.0pt\hbox{$\mathchar"13C$}}}
\def\gta{\mathrel{\spose{\lower 3pt\hbox{$\mathchar"218$}}
     \raise 2.0pt\hbox{$\mathchar"13E$}}}
\def\mincir{\ \raise -2.truept\hbox{\rlap{\hbox{$\sim$}}\raise5.truept  
\hbox{$<$}\ }}                                                          %
\def\magcir{\ \raise -2.truept\hbox{\rlap{\hbox{$\sim$}}\raise5.truept  %
\hbox{$>$}\ }}                                                          %
\def\simlt{\ \raise -2.truept\hbox{\rlap{\hbox{$\sim$}}\raise5.truept   
\hbox{$<$}\ }}                                                          %
\def\simgt{\ \raise -2.truept\hbox{\rlap{\hbox{$\sim$}}\raise5.truept   %
\hbox{$>$}\ }}                                                          %
\def\newline{\par\noindent}

\def\simleq{\; \raise0.3ex\hbox{$<$\kern-0.75em \raise-1.1ex\hbox{$\sim$}}\; }
\def\simgeq{\; \raise0.3ex\hbox{$>$\kern-0.75em \raise-1.1ex\hbox{$\sim$}}\; }
\newcommand{\eV}{{\rm eV}}
\newcommand{\keV}{{\rm keV}}
\newcommand{\GeV}{{\rm GeV}}
\newcommand{\TeV}{{\rm TeV}}
\newcommand{\Mpc}{{\rm Mpc}}
\newcommand{\kpc}{{\rm kpc}}
\newcommand{\muG}{\mu{\rm G}}

\begin{document}


\begin{center}
{\Large \bf Diffusion of  Ultra High Energy Protons in \\
Galaxy Clusters\\
and Secondary X and Gamma Ray Emissions}
\vskip1.cm
{\bf Corentin Rordorf$^{1,2}$, Dario~Grasso$^2$ and Klaus Dolag$^3$}  
\vskip0.5cm
$^1${\small\it  Ecole Polytechnique Federale de Lausanne, Switzerland}\\
$^2${\small\it Scuola Normale Superiore and I.N.F.N., Pisa, Italy}\\
$^3${\small\it Dipartimento di Astronomia, Universit\`a di Padova,
Padua, Italy}
\end{center}

\begin{abstract}
\noindent
In this work we simulate the propagation of Ultra High Energy (UHE) protons in the magnetised
intergalactic medium of Galaxy Clusters (GCs). Differently from previous works on the subject, we  
trace proton trajectories in configurations of the Intra Cluster Magnetic Field (ICMF)
which have been extracted from a constrained Magnetic-SPH simulation of the 
local universe.  Such an approach allows us to  take into account the effects of 
several  features of the ICMFs, e.g. irregular geometrical structure and field fluctuations due 
to merger shocks,
which cannot be investigated analitically or with usual numerical simulations. 
Furthermore, we are able to simulate a set of clusters 
which have  properties quite similar to those of GCs observed in the nearby universe. 
We estimate the time that UHE protons take to get out of the clusters
and found that in the energy range $5\times 10^{18} \simleq E \simleq 
3 \times 10^{19}~\eV$ proton propagation takes place in the Bohm scattering 
diffusion regime passing smoothly to a small pitch angle diffusion regime at larger energies.
 We apply our results to estimate the secondary gamma and Hard X Ray (HXR) emissions produced 
by UHE protons in a rich GC. We show that the main emission channel is due  
to the synchrotron HXR radiation of secondary electrons originated by proton photo-pair production
scattering onto the CMB. This process may give rise to a detectable
signal if a relatively powerful AGN, or a dead quasar,  accelerating protons at UHEs is harboured 
by a rich GC in the local universe.  
\end{abstract}
\vskip1.cm
PACS: 98.65Cw, 98.70Sa, 98.70Qy, 98.70Rz 
\newpage

\section{Introduction}

Several arguments  suggest that  Galaxy Clusters (GCs) may harbour sources of Ultra High Energy Cosmic 
Rays (UHECRs).  If UHECRs are produced by sources located in the galaxies, e.g. compact stellar remnants 
or relativistic shocks produced by stellar collapse, these sources will be abundant in GCs which contains 
hundreds or thousands of galaxies.    Active Galactic Nuclei (AGN), which are among the most promising 
source of UHECRs \cite{Berezinsky}, are typically harboured by elliptical galaxies which are more 
frequently observed in GC's.   
Furthermore, GCs  may act themselves as accelerators  of cosmic rays up to ultra high energies due to 
the presence of large scale magnetic fields and  shock waves produced during the cluster hierarchical 
accretion \cite{Bere97}.

 GCs are permeated by a magnetised  plasma of electrons and baryons.
 The density of the Intra Cluster Medium (ICM)   is determined by the observation of the X-ray 
bremsstrahlung  emission of  the gravitationally heated electron gas.  The inferred  density is  
 $n_{\rm gas} \sim 10^{-3}~\rm{cm}^{-3}$  which amounts to an over-density
 of about one thousand  with respect to the Inter Galactic Medium (IGM).  
  The presence of Intra Cluster Magnetic Fields (ICMFs) is testified by the extended radio halos due 
to the synchrotron emission of relativistic  electrons in the ICM. 
  The strength of the ICMFs  can be estimated from the intensity of the observed radio emission
either assuming the minimum energy condition, giving $<B> \sim 0.1 - 1~\muG$ \cite{Feretti99}
($<B>  \sim 0.4 ~\muG$ for Coma \cite{Giovannini93})
 or  by an  independent determination of the relativistic electrons density.  For a few clusters 
 (especially for Coma)  this is made possible by the observation of a HXR emission that,  if 
 interpreted  as the outcome of Inverse Compton Scattering (ICS) of the relativistic electrons onto 
the CMB photons, implies the ICMF  strength to be in the range $0.2 - 0.4 \muG$ \cite{Fusco,Rephaeli99}  
However, the interpretation of the HXR radiation from GCs in terms of ICS  is  still controversial
and other models have been proposed which may allow/require stronger ICMF
(see e.g. \cite{Blasi99,Atoyan99,Neronov03}) . 
 Indeed,  Faraday Rotation Measurements (RMs) of polarised radio sources placed within 
the cluster, or in the background,  provide  significant evidence for the presence of stronger ICMFs 
 in the range $1 - 10~\muG$ or higher \cite{Feretti99b,Taylor01}. 
Furthermore,  RMs  provide valuable information about the  spatial structure of the field
which is patchy with coherence lengths in the range  between 10 and  100 kpc. 
It is unclear if the discrepancy between the IGMF strength inferred from the radio halo intensity and 
that determined from RMs can be explained by the different sensitivity of these two kind
of measurements to the presence of magnetic substructures (for a  discussion about this issue
and a comprehensive review of IGMF observations see   \cite{Carilli02}). 
Interestingly, a recent analysis \cite{Fusco03} of BeppoSAX data on the HXR emission from Coma
points to a ICMF strength in that cluster  which is stronger  ($<B> \sim 1~\muG$) than the value 
previously claimed in \cite{Fusco} and it is almost compatible with that inferred from RMs. 
In the following we will assume that the actual strength of IGMFs is that inferred from RMs.
 
The magnetised ICM can affect significantly the propagation of cosmic rays up to ultra high energies.  
The Larmor radius of protons  in the ICMFs is  smaller than the field coherence length up
ultra high energies so that the residence time in a GC can be considerably increased respect to
the straight propagation. The effect is even more pronounced for 
composite nuclei which have a smaller Larmor radius. Here we will focus on the case of protons.
At low energies the propagation UHE protons is  expected to take place in the spatial 
diffusion regime.  Since diffusion arises due to the scattering of charged particles onto the field 
irregularities, the diffusion time is expected to depend  significantly on the ICMF power spectrum.   
Although very interesting approaches have been recently proposed to
determine the ICMF power spectrum from high resolution RMs \cite{Vogt03, Murgia04},  
so far this is  a poorly known quantity.
The common attitude is to assume a Kolmogorov power spectrum.  Although this is the most natural choice 
in the case of fully developed,  homogeneous turbulence of  Navier-Stokes type (no-magnetic) we should 
take in mind that some, or any,  of these conditions may not be fulfilled  in the ICM.   Indeed, the ICMF
 power spectrum may  depend on the cluster  accretion history.   Furthermore,  cluster merging is
  expected to compress and twist locally the ICMF giving rise to  non-Gaussian MF fluctuations
  which are  non accounted for in conventional CR diffusion simulations which use a  synthetic 
  MF with  Gaussian fluctuations.
 The only way to account for  the effects of all  these  ICMF features 
 on the propagation of UHECRs  is by means of numerical simulations of the MF large scale structure. 
Numerical simulations are also mandatory to simulate CR diffusion when the Larmor radius 
is of the same order of  the MF coherence length.  We will show that this is the case for
protons with energy around $10^{19}~\eV$, i.e. the most interesting energy range
for the extensive air shower (EAS) experiments.
Furthermore the determination of the diffusion coefficients can be performed 
analytically only for MF in a regime of weak turbulence.   We will show that this may not be the 
case for ICMFs. 
  
  The main aim of this paper is to estimate the diffusion time of UHE protons in rich GCs
as a function of their energy.  
  Differently from previous works  on the subject we approach this problem  by means of numerical 
simulations to determine  both the ICMF  structure of nearby GCs and to trace proton propagation. 
We extract the ICMF in several GCs from a constrained simulation of the MF structure in
 the local universe \cite{Dolag03}   based on  the Magnetic Smoothed Particle Hydrodynamics (MSPH) 
technique developed by Dolag et al. \cite{Dolag02}.  
 This simulation traces the passive evolution of the magnetic field in a spherical region of radius 
115 Mpc about the Local Group  starting from a primordial seed field.  
It was showed  \cite{Dolag03,Dolag02} that the simulation  reproduces  the observed RMs  
with good accuracy.  The main advantage of this approach is that it provides a realistic
simulation of the gas and of magnetic field spatial distribution in a number of nearby clusters.
It was found that on average the ICMF  traces the gas density and it is locally amplified in regions
where accretion shocks fronts are present.     
 The ICMF power spectrum has been  determined down to the spatial resolution length
 which is $\sim 10~\kpc$ in the cluster center.  Significant deviations were found 
in some GCs respect to a Kolmogorov spectrum.  
 The effects produced by  all these features  of the ICMF  onto the UHECR  diffusion  can be relevant 
and were not considered in the existing literature. Our approach allows  to take them into account.
 
As an application of our results we estimate the flux of  HXR and gamma rays produced by the 
interaction of UHE protons with the gas and the radiation within GCs.  
We will mainly focus on the  synchrotron emission of ultra-relativistic secondary electrons, produced by $pp$-scattering, which falls in the GeV region, and by  proton photo-pair scattering onto the CMB falling in the HXR region of the electromagnetic spectrum.
Whereas the former emission can hardly be detectable, the latter may be detectable if 
a relatively bright UHE proton source is harboured by a rich GC in the local supercluster.
 
In Sec. 2 we give a brief description of the MSPH simulation and of the  ICMF properties
obtained from this simulation.  
In Sec. 3 we describe our ray-tracing code and  the UHE proton diffusion properties
 at different energies and radii. 
In Sec. 4 we show that the insertion of MHD turbulence at low length scales, which is unresolved
in the MSPH simulation, does not affect significantly our results.  In Sec. 5 we compare our approach 
with that  followed by other authors.
In Sec. 6  we estimate the secondary gamma-ray emission  due to hadronic scattering onto the
intra-cluster gas and the HXR radiation due to the synchrotron emission of electrons and positrons 
generated by proton photo-pair production scattering. Finally, Sec. 7 contains our conclusions.

\section{MSPH simulations and ICMF properties}

The origin of  ICMFs is still unknown. 
Although the observed high metallicity of the ICM suggests that it may have undergone
a significant pollution  driven by galactic winds , the strong intensity of ICMFs, their   
huge extension (exceeding  few Mpc's in some cases), and their large coherence length 
($10 \div 100~\kpc$'s) make hard  to  explain their origin uniquely by galactic ejection. 
 Another possibility is that ICMFs were generated starting from a  seed  which is subsequently amplified
  by the adiabatic compression and the shear flows  of the gas driven by the hierarchical accretion of 
the clusters.
  Several mechanism have been proposed to explain the origin of the required seed field. 
For example, it could be ejected by AGN \cite{Furlanetto}, 
 or from a violent starburst activity at high redshift \cite{Volk98}, 
it  could be produced by some non-equilibrium process in the early universe \cite{report}, or be  the 
result of a Biermann battery \cite{battery}. 
 In principle, the battery has the advantage to be independent on  unknown  physics at high 
  redshift since, in this case, the seed field is produced by thermoelectric currents powered by the 
cluster merger shocks.
In practice, however,  the battery can account only for too weak  magnetic seeds \cite{Kulsrud97}.  
A subsequent turbulent dynamo has to be invoked to increase the  field strength enough to initiate  
a successful  MHD amplification.  Since  this process cannot be simulated on the computer, the final 
intensity  of the magnetic field is  quite uncertain and has to be tuned 
to reproduce the RMs of GCs.
Operatively, such an approach is, therefore, similar to assume a seed field of primordial origin.     
   
 The simulation that we use here is based on the  MSPH method developed in \cite{Dolag99}.  
The code combines the merely gravitational interaction of the dominant dark-matter component
with the MHD of the electron-baryon gas.   
 In a previous work \cite{Dolag02} it was showed that magnetic seed
fields in the range of $(0.2-5) \times 10^{-9}\,{\rm G}$ at redshift
$z_* \simeq 20$ will be amplified due to the structure formation
process and reproduce RM in clusters of galaxies. This corresponds to
$B_0 \equiv B(z_{*}) (1 + z_{*})^{-2} \simeq (0.05-1) \times 10^{-11}$ G at the present
time in the unclustered IGM. It was also
demonstrated that the MF amplification process completely erases any
memory of the initial field configuration in high density regions like
galaxy clusters.  Therefore, we can safely set the coherence length
$L_c(z_{\rm in})$ of the initial seed field to be infinite in our
simulation.  
    
In the simulation which we use here $B_0 = 10^{-12}$ Gauss. The initial conditions for the DM 
fluctuations were constructed from the IRAS 1.2-Jy galaxy survey by first smoothing the observed
galaxy density field on a scale of 7 Mpc, evolving it linearly back in time, and then using it as a 
Gaussian constraint for an otherwise random realization of the $\Lambda$CDM cosmology. We extended the initial conditions of~\cite{Mathis} by adding gas, together with an initial MF. 
Therefore the simulation at redshift zero represents the large scale mass distribution of our local
universe as observed. Some of the prominent halos in the simulations can be identified with observed
counterparts with mass and temperature within a factor of two or better. In \cite{Mathis}
 it is shown, that this simulation matches the observed local universe very well in many aspects. 
Therefore, such an approach allows to use a simulated set of GCs which have very similar 
properties to those of  GCs observed nearby. 

The gravitational force resolution of the simulation is $10\,{\rm kpc}$. This maximal resolution is
reached in the central region of GCs where the SPH test particles are more dense. 
The volume filled by the mixture 
of high resolution dark matter and gas particles is a sphere of radius $\sim 115$ Mpc
(centered on the Milky Way) surrounded by a larger region of low resolution dark matter particles,
 to get long range tidal forces. The mass of the high resolution gas and dark matter particles
thereby is $0.48 \times 10^9M_\odot/h$ and $3.1 \times 10^9M_\odot/h$ respectively. Therefore the most 
massive cluster in our simulation is resolved by nearly one million particles.

\begin{figure}[h!]
\centerline{\mbox{\psfig{figure=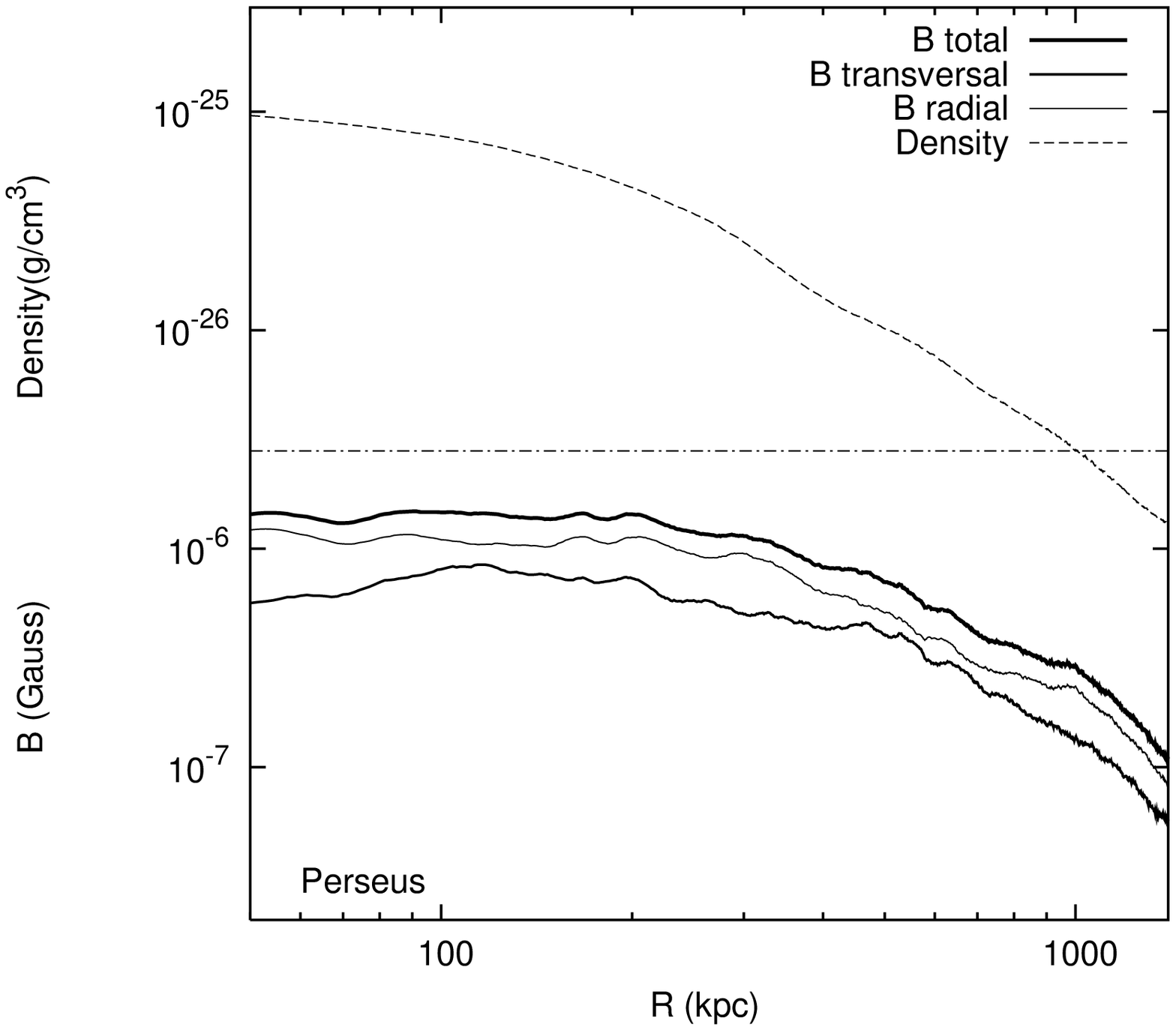, angle =0, width = 7 cm}
\psfig{figure=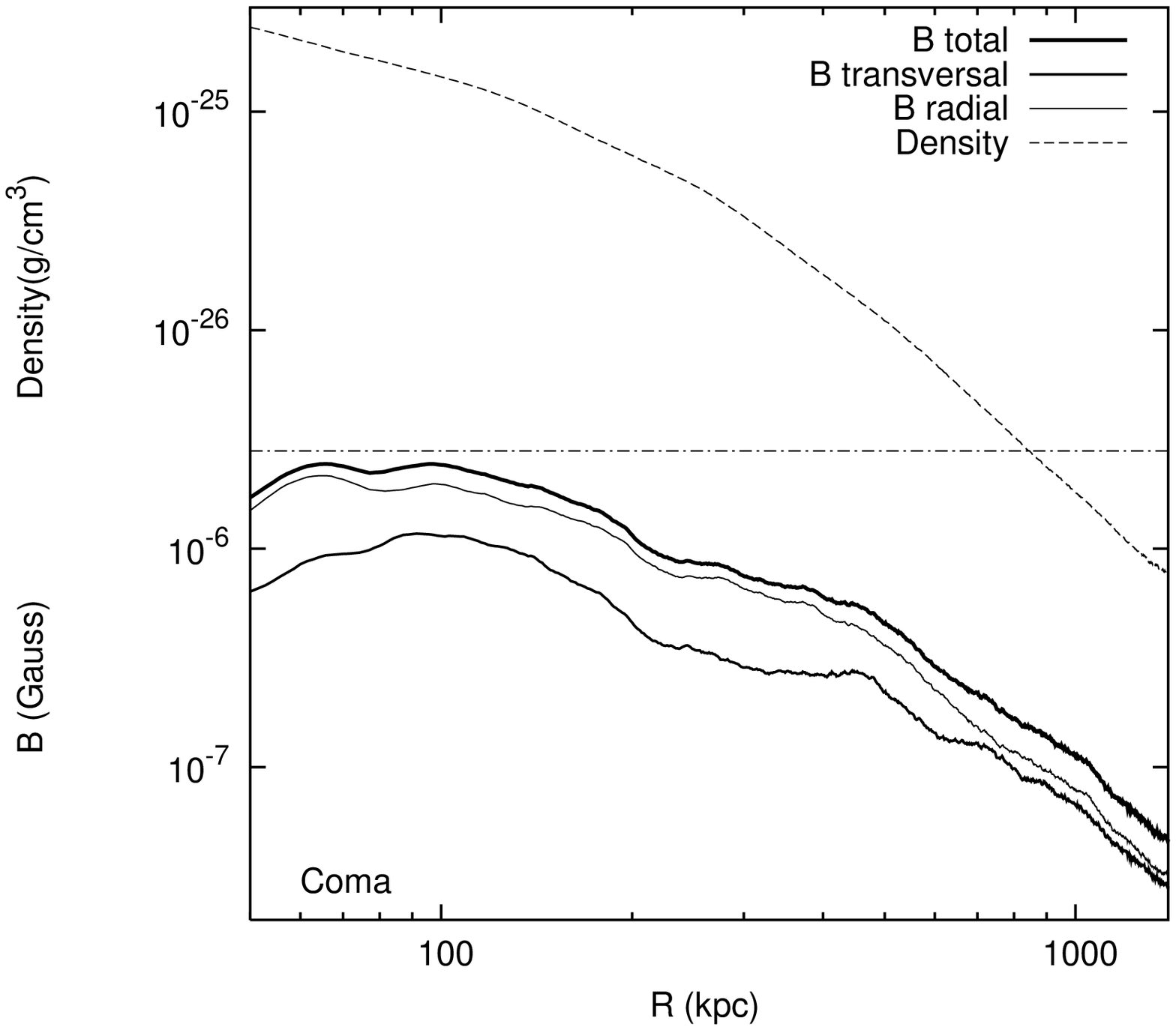, angle =0, width = 7 cm}}}
\caption{Density and magnetic field radial profiles of clusters 2 and  4.
Different  spatial components of the ICMF are represented.}
\label{prof}
\end{figure}

For our analysis we used the four most massive clusters within our simulation volume, which 
includes the halos identified as the Perseus and the Coma cluster. Table \ref{Tab:1} summarizes the
global properties of these clusters.
\begin{table}[h!]
\begin{center}
\caption{Cluster number, virial radii, virial mass, emission weighted virial temperature 
and identification with real clusters (if possible).} 
\label{Table}
\begin{tabular}{cccccc}
\#&$R_{200} [\kpc/h]$&$M_{200} [M_\odot/h$] & $T_{\rm Lx} [{\rm keV}]$
&name\\
\hline \hline
1& 2421.76&1.60e+15& 7.1 & - &\\
2& 2257.75 & 1.30e+15& 5.6 & Perseus \\
3& 1958.90& 8.50e+14 & 4.6 & - &\\
4& 1885.23& 7.57e+14& 8.3 & Coma \\
\end{tabular} 
\label{Tab:1}
\end{center}
\end{table}
We determined the gas density and magnetic field profiles for the clusters. On average, the magnetic 
field profiles follow the density profile in the outer part, whereas in the central region
the magnetic field profiles flatten, which is consistent with previous findings. Depending on the 
cluster and its dynamical state, the slope of the magnetic profile in the outer parts
scatters around the slope of the gas density, and also the size of the flatter core region varies. 
Fig. \ref{prof} shows, as an  example, the profiles for the halo 2 and the halo 4. For the magnetic
field we present the norm as well as the radial and transversal component. 
It is evident from these plots that the MF structure is basically isotropic.

We also determined the MF power spectrum $B^2(k)$. As we will discuss in more details in the 
Sec.3, this quantity is crucial to determine CR diffusion properties in the ICM. We
computed $B^2(k)$ by evaluating the MF at the points of a cubic grid. This is done by applying the
MSPH formalism to every mesh point of the grid considering all particles which overlap by their
smoothing length, taken from the MSPH simulation. Afterwards we performed a FFT (Fast
Fourier Transform) on this grid.

\begin{figure}[h!]
\centerline{\mbox{\psfig{figure=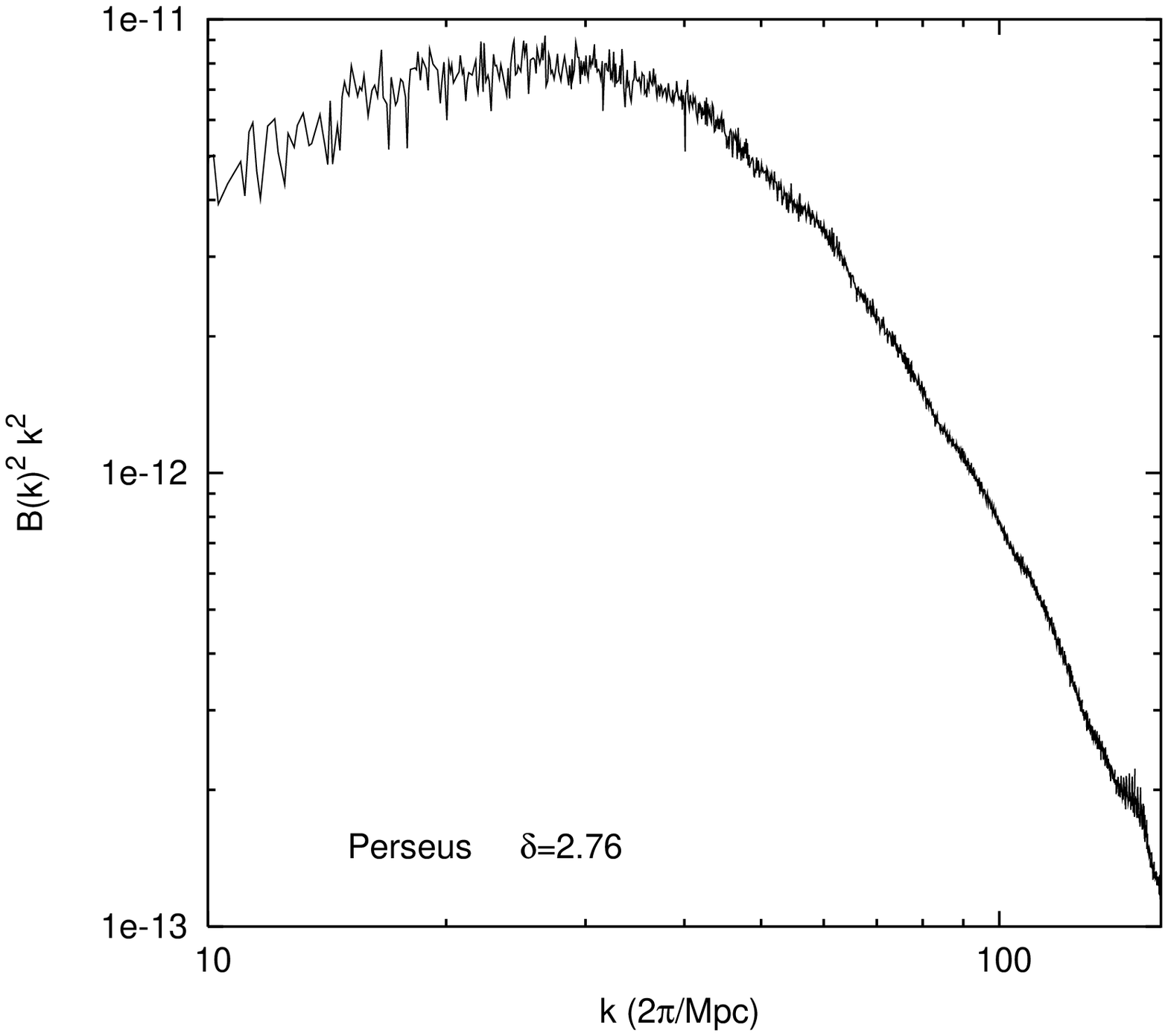, angle =0, width = 7 cm}
\psfig{figure=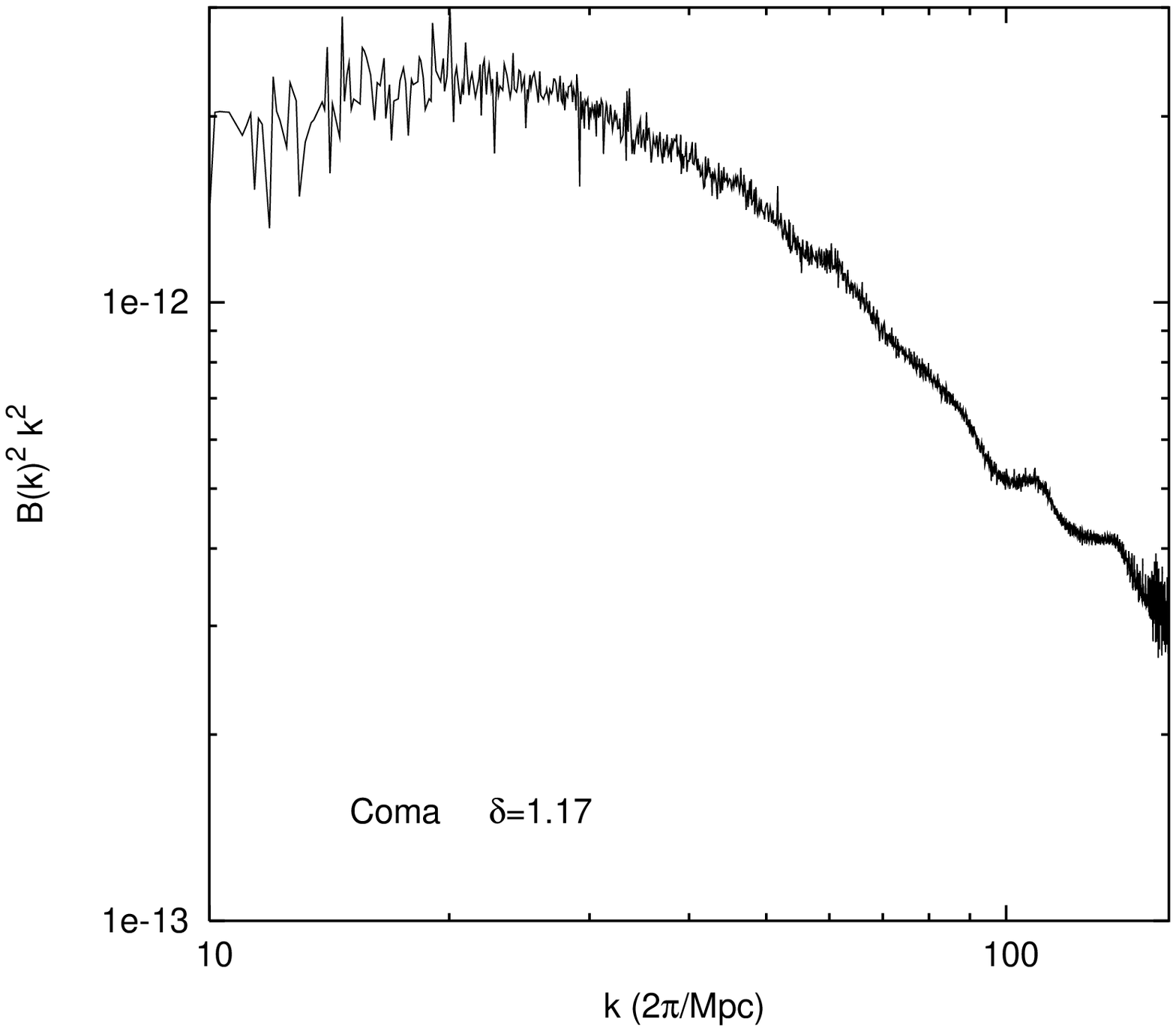, angle =0, width = 7 cm}}}
\caption{The energy power spectrum is represented for  clusters  2 (left) and 4 (right)
in arbitrary units.  The quoted values of  $\delta$ represent the best fit to the  
 power law index. $|\delta| = 5/3$ for a Kolmogorov spectrum. }
\label{pws}
\end{figure}

In Fig.\ref{pws} we show the energy spectrum ($B(k)^2 k^2$) for clusters 2 and 4. The slope of 
the power spectrum for the four most massive haloes seems to be steeper for the more massive 
ones, but when calculating the spectra for all clusters it looks more
likely that it is determined by the dynamical state of the cluster. 
The spectral index $\delta$  ranges between -1.5 and -3 for 
most of the clusters, with a small number of clusters found to have even more extreme values.
On average we found $\delta \simeq -2$ which corresponds to a spectrum slightly steeper 
 than a Kolmogorov's for which (in 3D) $\delta = - 5/3$.  This is 
consistent with the results obtained in the first attempts to measure the
ICMF power spectra from RM images of extended radio sources 
(\cite{Vogt03,Murgia04}).

In order to quantify the turbulence strength we define the parameter
\begin{equation}
\eta = \frac{ \langle \delta B^2\rangle}{ \langle B^2\rangle}  = 
\frac{\frac{1}{2\pi^3}\int_{k_{\rm min}}^{k_{\rm max}} dk k^2 B^2(k)}
{\int B^2({\bf x})d^3x}
\end{equation}
where $k_{\rm min}$ is the wavenumber at which the energy spectrum gets its maximal value and
the turbulent cascade sets-in. This is the scale at which most of the MF energy is concentrated.
 $k_{\rm max}$ is determined by the spatial resolution of the MSPH simulation. Although the
 physical wavenumber at which dissipation takes place is expected to be much larger, this artificial 
cut will not affect significantly $\eta$ due to the rapid decreasing of the turbulent power at small 
scales.

From the Figs.\ref{pws} we can estimate $k_{\rm min} \simeq 50~\frac{2\pi}{\kpc}$ almost 
independently on the cluster mass. We have then a turbulence strength $\eta \simeq 0.6$ over 1 
Mpc box. In other words, our MSPH simulation predicts a quite high level of turbulence in rich GCs.
      
\section{Ray tracing and UHE proton distribution in galaxy clusters}

We simulate proton trajectories in the synthetic ICMF by solving equation of motion 
by means of a Runge-Kutta adaptive step size method.  Our simulation conserves energy with a very good 
accuracy: we get $\frac{\delta E}{E} \simeq 10^{-5}$ in the worst case of  $10^{18}~\eV$ protons over a 
complete trajectory. As far as it concerns the simulation of proton trajectories, we  disregard energy 
losses.   
We will verify the validity of this assumption  {\it a posteriori} by comparing the residence time in 
clusters with the energy loss time due to the relevant collision processes.  
The ICMF is determined  along the proton trajectory by performing a  weighted sum 
(see \cite{Dolag99} for details) of the magnetic fields of all smoothed gas particles  in the MSPH 
simulation which overlap  with the proton position.  In same cases an extra turbulent MF 
component is artificially added  to the result of the simulation to model  a possible unresolved,
 or physically unaccounted,  small scale component of the IGMF (see Sec. 5). 
 
\begin{figure}[h!]
\centerline{\epsfxsize=9cm \epsfbox{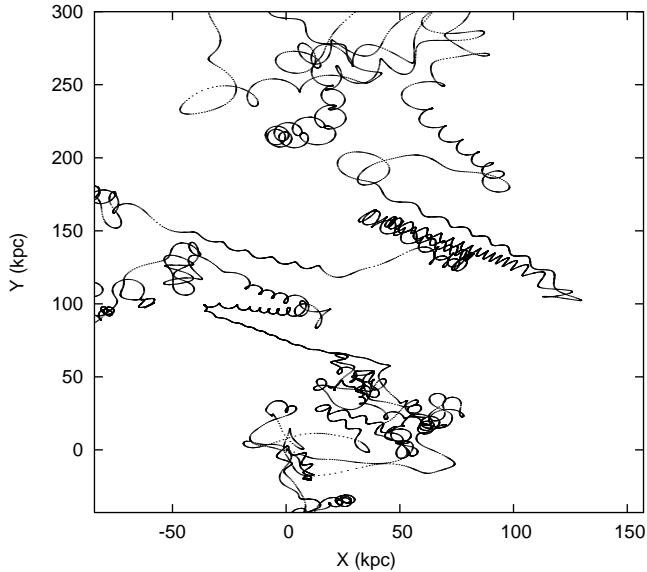}}
\caption{Trajectories of a $E = 5 \times 10^{18}~\eV$ proton in the central region of 
cluster 1 projected in 2D. } 
\label{2dtrak}
\end{figure}

\begin{figure}[h!]
\centerline{\epsfxsize=9cm \epsfbox{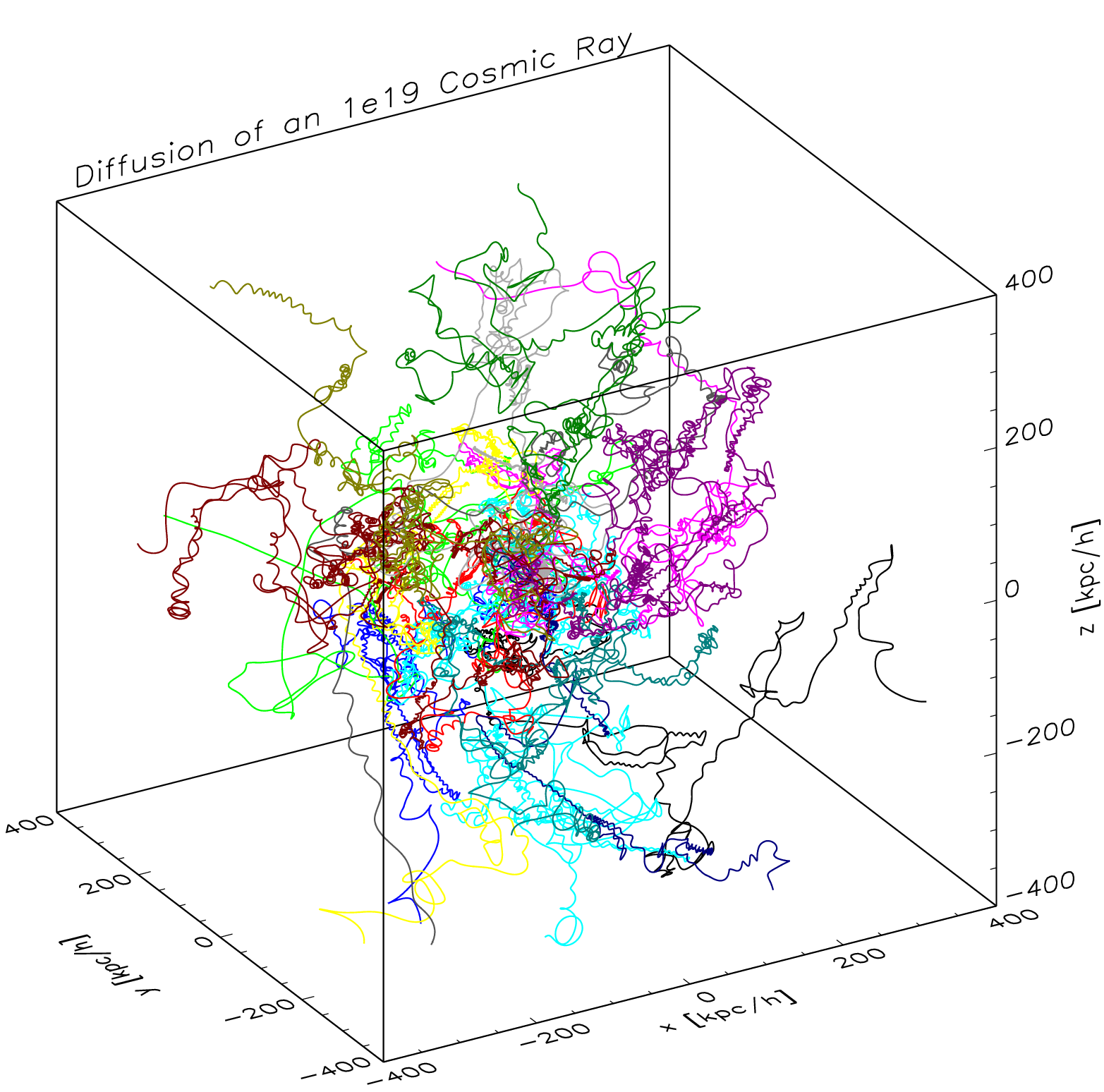}}
\caption{The same of Fig.\ref{2dtrak} in 3D and for protons with energy $10^{19}~\eV$. }
\label{3dtrak}
\end{figure}

In  Figs.(\ref{2dtrak},\ref{3dtrak})  we represent typical proton trajectories
in two and three dimensions.
These plots  clearly show that in the central regions of a rich GC  proton propagation takes place in the spatial diffusion regime at energies as large as $\sim 10^{19}~\eV$.  Several  momentum reversals are visible which we interpret to be due to magnetic mirroring onto local enhancements of the magnetic field strength.  


\begin{figure}[h!]
\centerline{\epsfxsize=11cm \epsfbox{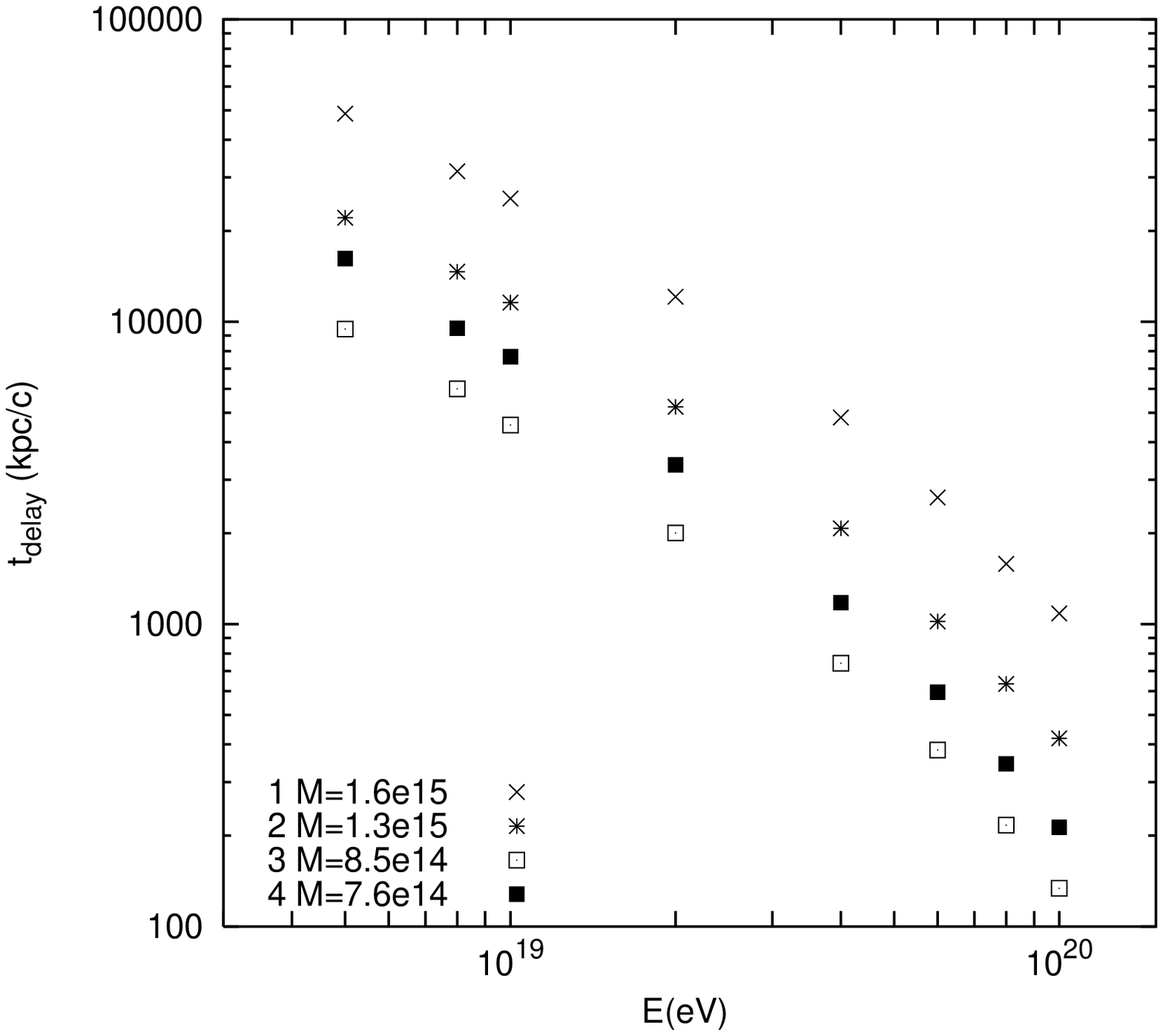}}
\caption{The delay time  (residence time within a sphere of radius $0.5~\Mpc$ minus the 
straight propagation time) for 4 GCs. The quoted GC masses are in solar masse units. }
\label{tdelay}
\end{figure}

We start  computing the mean delay time, i.e. the effective arrival time minus the straight 
propagation time to reach a sphere far out of the cluster (we take it  at $35~\Mpc$ radial 
distance), of 1000 protons injected at the cluster center   with  energies in the range 
$ 10^{18}-10^{20}~\eV$.   
In Fig.\ref{tdelay} we represent the delay time as a function of the proton energy for several GCs.
The best fit function is well represented by two power laws with  indexes  $\sim -1$ and $\sim -2$,
the  knee being located  at an energy  $E_{\rm cr} \sim (2 - 3) \times 10^{19}~\eV$. 
In Sec.5  we will argue how this feature corresponds to the protons propagation passing  from  the 
spatial diffusion  (SD) regime to the small pitch angle (SPA) regime at  high energies. 

\begin{figure}[h!]
\centerline{\mbox{\psfig{figure=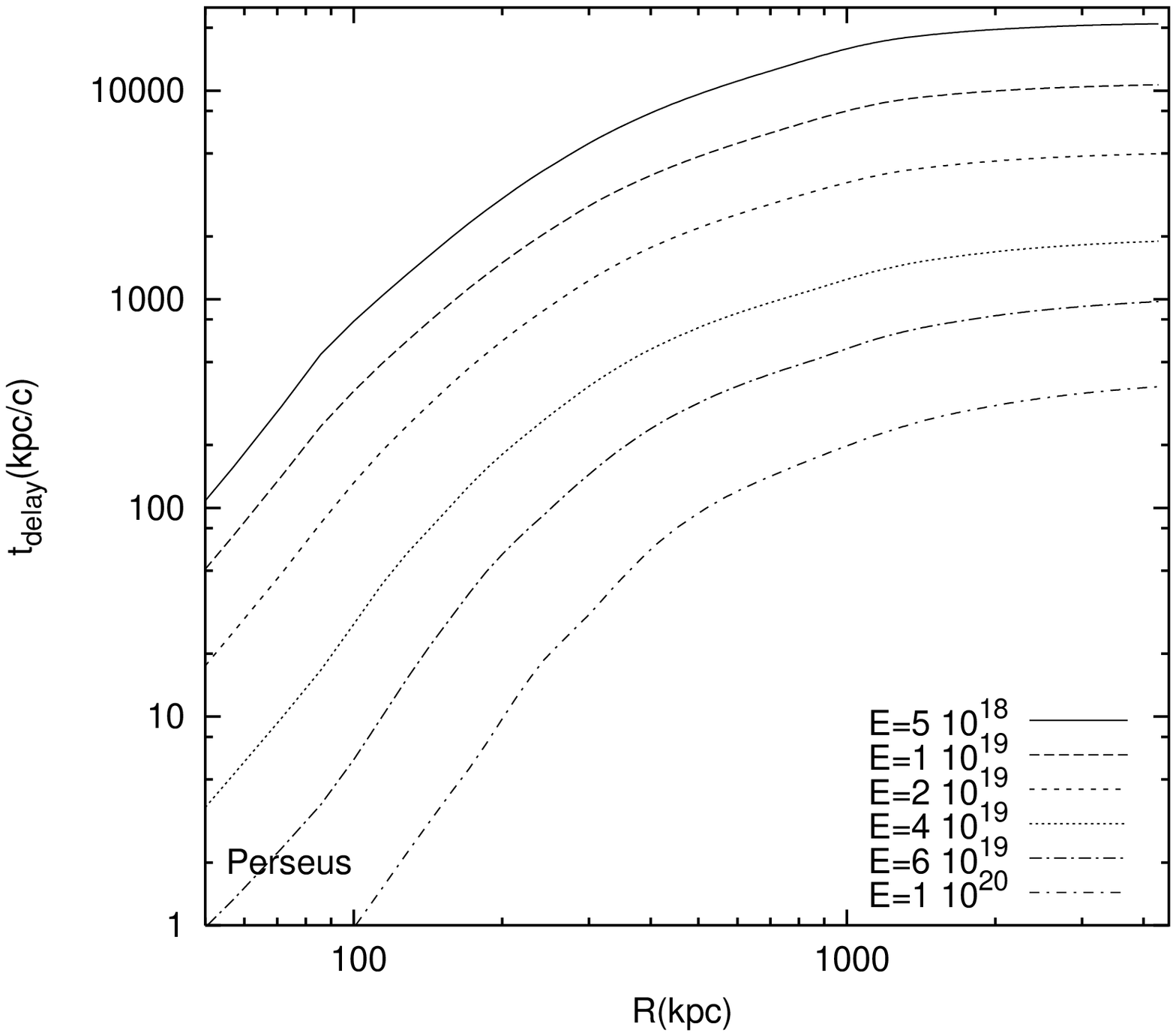, angle =0, width = 7 cm}
\psfig{figure=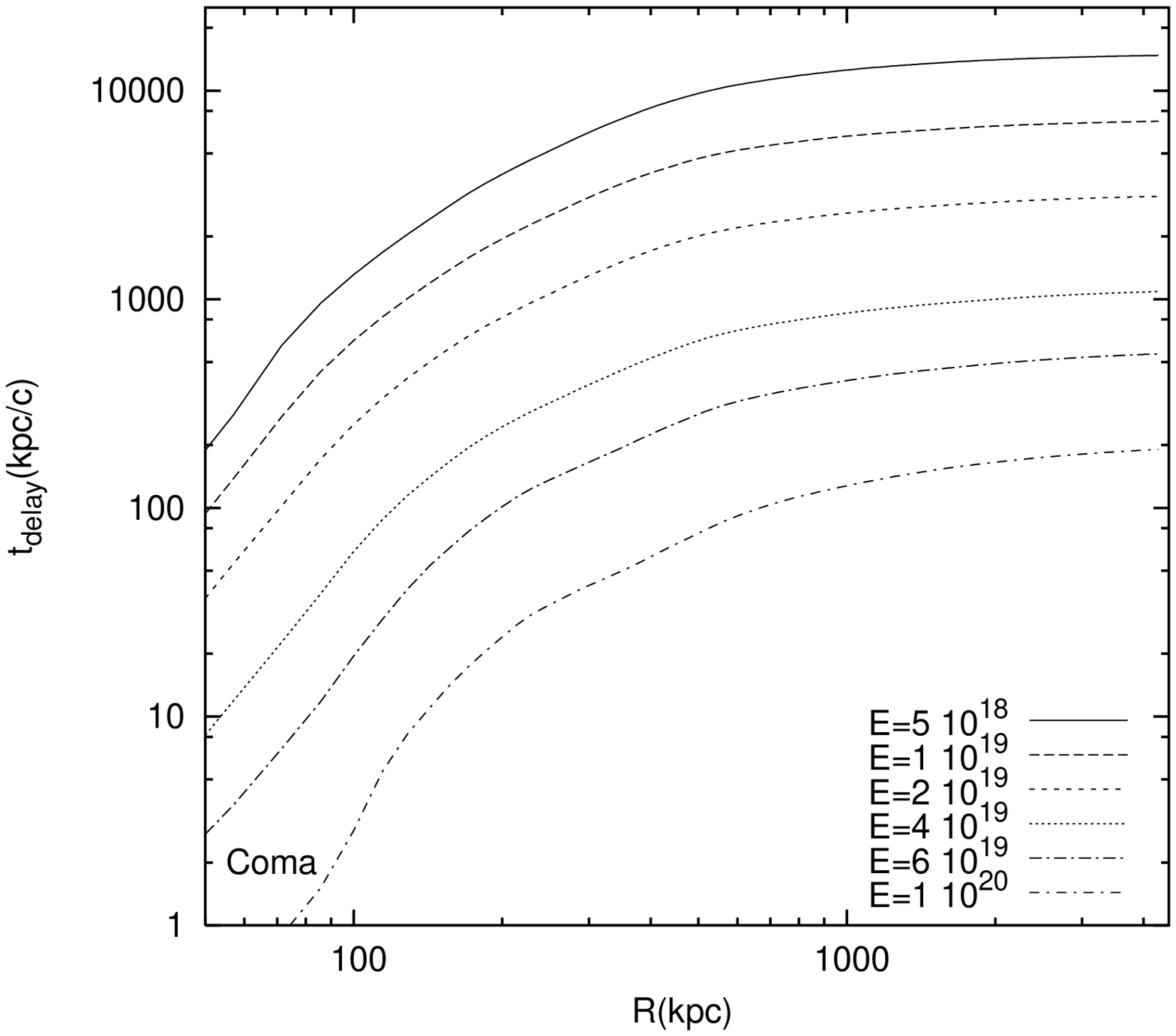, angle =0, width = 7 cm}}} 
\caption{The mean delay time accumulated from the cluster center to the radius R is represented 
as a function of R for protons of different energies (quoted in eV's)  for clusters 2 (left) and 
4 (right).}
\label{tdelay_R}
\end{figure}

The  delay time, that we defined in the above, is representative of the diffusive properties of the 
entire cluster. In order to get an insight on the radial dependence of this quantity, we  computed
 the delay time from the center up to spheres of different radii. This is represented in 
the Figs. \ref{tdelay_R} at several energies for clusters 2 and 4.    
These plots show that no significant  delay is produced at a radius larger than  1 Mpc . Interestingly, this feature is almost  independent on the proton energy as it is  apparent from 
the self-similarity of  the curves.  

Our next step is to compute a set of  time intervals $t_{ij}$ representing the mean time that a 
proton  injected in the spherical shell  $R_i < R < R_{i+1}$ spend in the  shell 
$R_j < R < R_{j+1}$ . Once we specified the injection spectrum and the radial distribution of 
sources, the knowledge of  a sufficiently dense set of $t_{ij}~'{\rm s}$ allows to determine the 
differential  proton flux at different radii: 
\begin{equation}
\frac{dF(E,R_i)}{dE}= Q_s(E) \frac{c}{4\pi}\sum_{j=1}^{m} \,
\frac{ (R_{j+1}^3 -R_{j}^3) } { (R_{i+1}^3 -R_{i}^3) } \,
 f\left({R_j}\right) ~t_{ij}~.
\label{profile}
\end{equation}  
Here $f(R_i)$ represents the density of sources and $Q_s(E)$ is the proton injection spectrum of a 
single source (for the sake of simplicity we assume all the sources to be identical).   
We took $m = 2000$ corresponding to as many identical shells  each of them is 1 kpc thick.  
The normalization  constant   $A$   depends on the source injection power. Eq. \ref{profile} 
amounts to a numerical solution of the diffusion equation. 

\begin{figure}[h!]
\centerline{\mbox{\psfig{figure=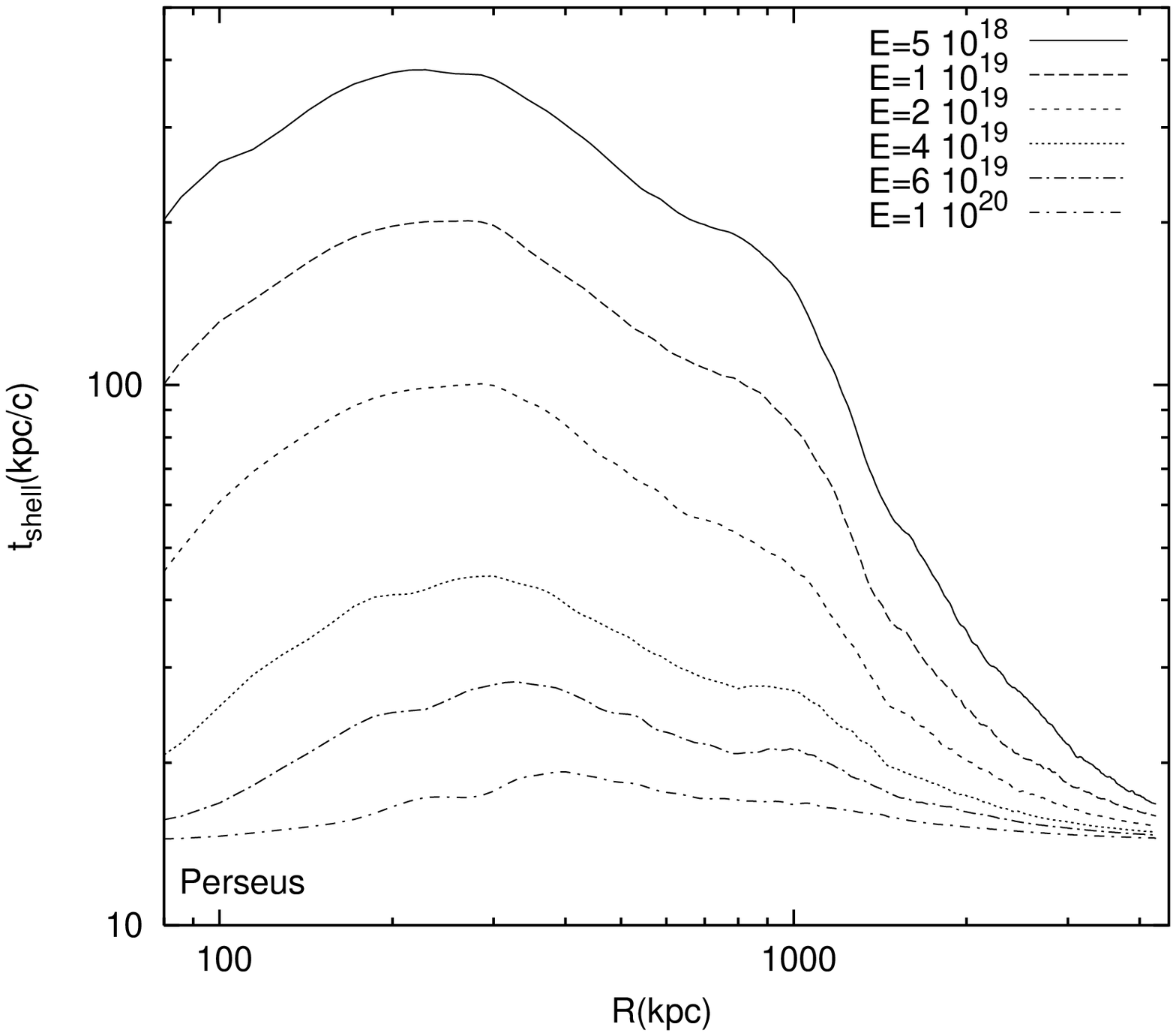, angle =0, width = 7 cm}
\psfig{figure=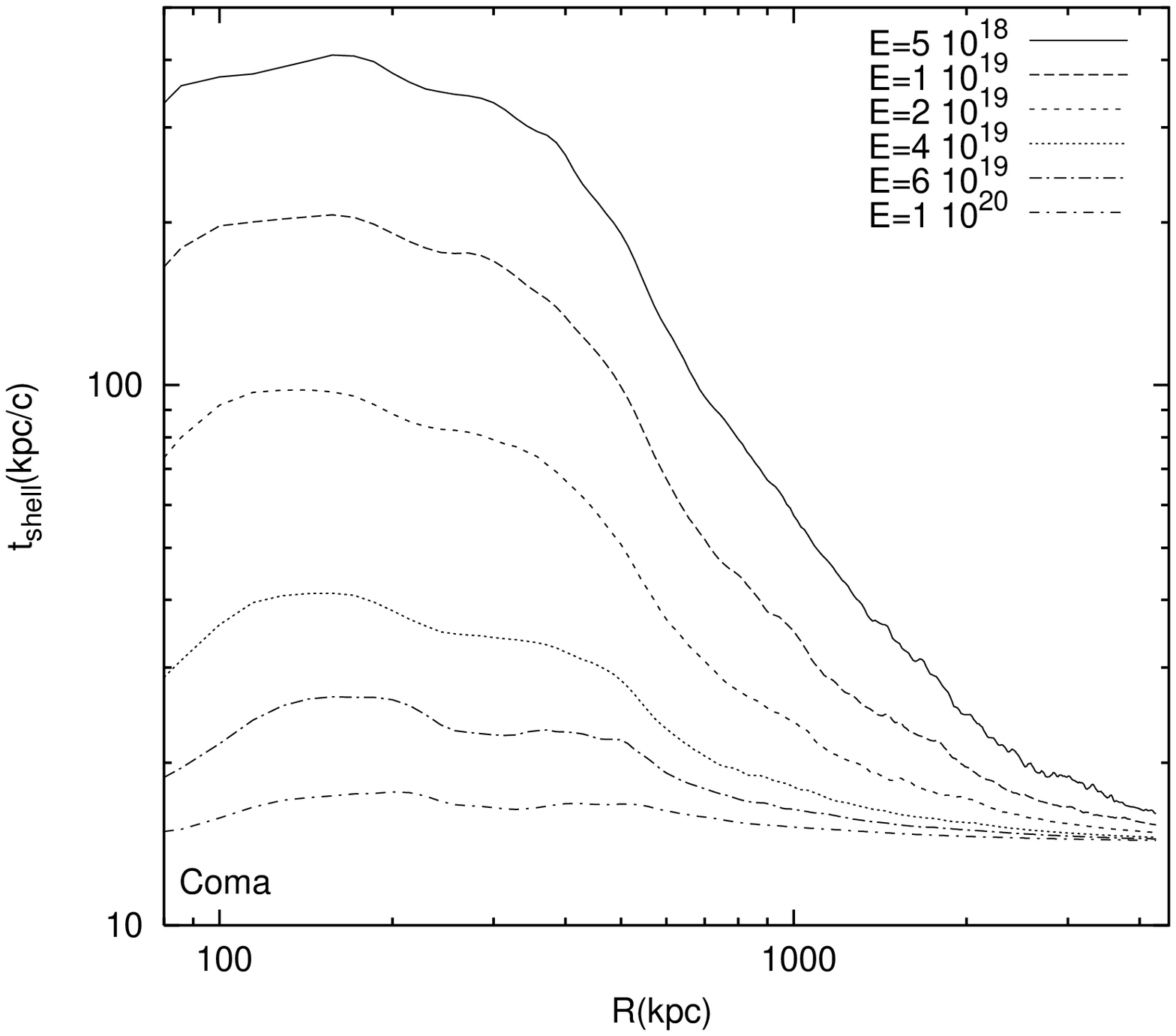, angle =0, width = 7 cm}}} 
\caption{The mean delay time accumulated in spherical shells of thickness 1 kpc and mean 
radius R for clusters 2 (left) and 4 (right).}
\label{t1R}
\end{figure}

In the case of a point-like single source placed at the center of the cluster,  
$f(R_i) = \delta_{1j} $, where $\delta_{ij} $ is the Kronecker function.     
It is interesting to plot the histogram of $t_{1j}$. For a better readability we represent it in 
Fig.\ref{t1R} in the continuous limit.

\begin{figure}[h!]
\centerline{\epsfxsize=10cm \epsfbox{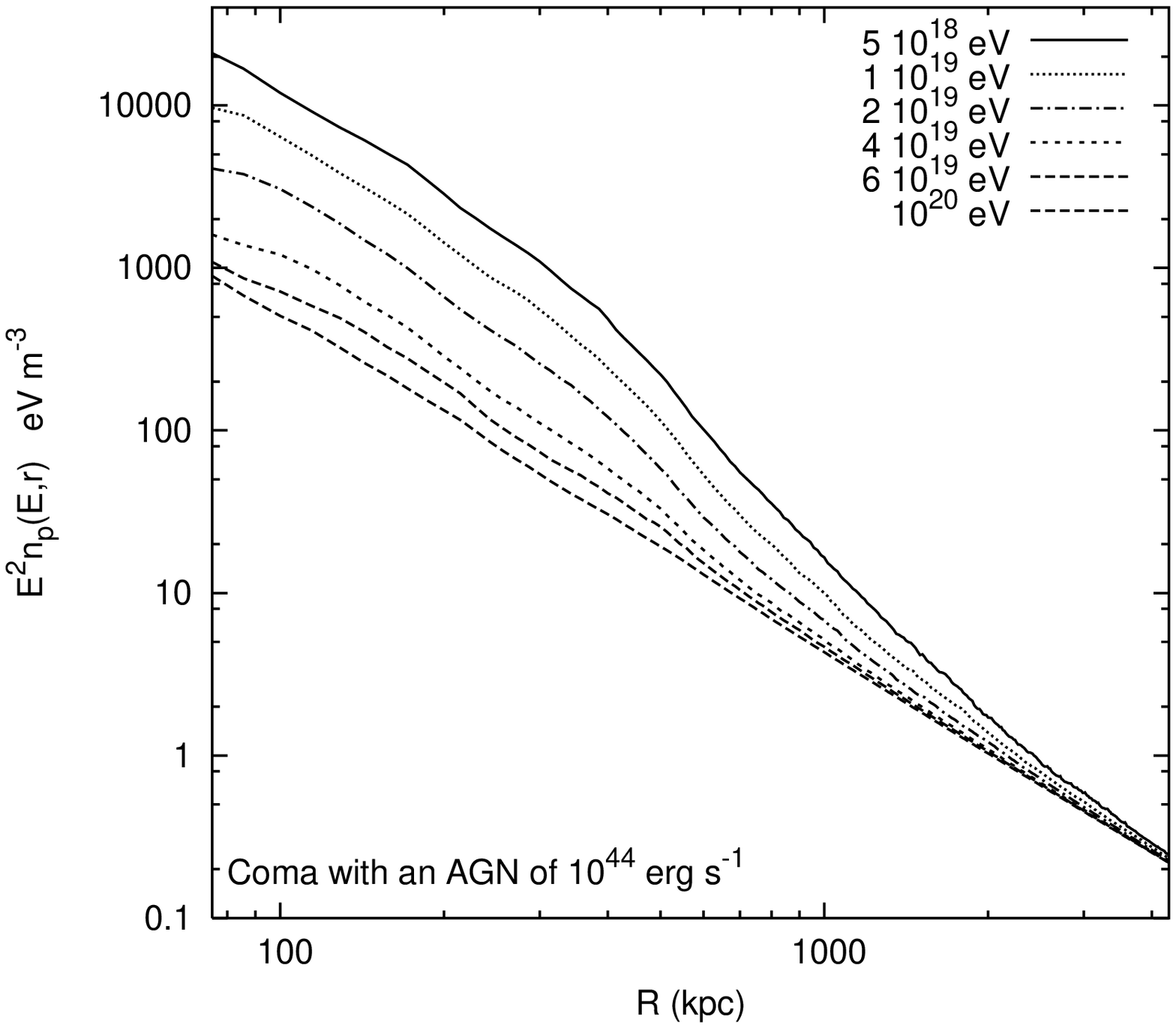}}
\caption{The mean proton energy density in the cluster 4 (Coma) is plotted
versus the radial distance from the center for several energies.}   
\label{AGN}
\end{figure}

In Fig.\ref{AGN}  we represent the  radial dependence of the proton  energy density per 
logarithmic energy interval at several energies in the case of a single source,
 with  luminosity  $L_p(E > 10^{18}~\eV) =   10^{44} {\rm erg~s}^{-1}$  and spectral 
index $\alpha = 2$   placed at the center of cluster 4 (Coma). 
 
Another  possibility is that  the UHECR source distribution  traces  that of galaxies.
We can reasonably assume that the radial distribution of galaxies is well approximated by
that of the gas.  To be consistent we determine the gas distribution from the result of the
MSPH simulation. We found that  
\begin{equation}
f(r)=\frac{1}{\left[1+\left(\frac{r}{r_1}\right)^2\right]^{0.51}}
\frac{1}{\left[ 1+\left(\frac{r}{r_2}\right)^2\right]^{0.72}}
\frac{1}{\left[1+\left(\frac{r}{r_3}\right)^2\right]^{0.58}}
\end{equation}
with $r_1 \simeq 10\kpc$ (spatial resolution limit), $r_2 \simeq 250~\kpc$
(core radius), $r_3 \simeq 2~\Mpc$ (virial radius) 
provides an excellent fit of the simulated gas distribution for  GCs with mass
$M \sim 10^{15}~M_\odot$. The mean UHECR luminosity of galaxies can be estimated by
 assuming that the Milky Way provides a representative sample. 
The high energy cosmic ray luminosity of our Galaxy have been estimated by several authors,
see e.g. \cite{Berezinsky} where it was found (we assume here that most of the UHECRs are protons
$ L^{\rm gal}_p(E > 5 \times 10^{18}~\eV) \sim 5 \times 10^{36}~{\rm erg/s}$.
Therefore, if the entire  luminosity of a rich  GC  is due to  $\sim 1000$ galaxies, 
its expected value would be in the range  $L_p(E > 5 \times 10^{18}~\eV) \simeq
10^{39} - 10^{40}~{\rm erg/s}$. We will show in Sec.6 that this luminosity is too low to give
rise to a detectable secondary emission from the ICM.  

A similar situation is expected if Gamma Ray Bursts (GRBs) are the main source of UHECRs. 
One of the  main arguments in favour of this hypothesis \cite{Waxman,Vietri} is the coincidence 
of the predicted UHECR flux  with that observed  under the assumption that the  energy 
release  under  the form of UHECRs  by GRBs equals that  of gamma rays.
The rate of GRBs in a  typical galaxy is $\Gamma_{\rm GRB}^{\rm gal} \sim 10^{-8}~{\rm year}^{-1}$,
hence  it is $\Gamma_{\rm GRB}^{\rm GC} \sim 10^{-5}~{\rm year}^{-1}$ in a  rich GC. 
Our previous results imply that  the diffusion time of UHE nuclei emitted  in the central region 
of a  rich cluster ($R \simleq 1~\Mpc$)  exceed  largely the  time interval between GRBs in the cluster.
As a consequence,  clusters behaves as continuous sources of UHECRs even if elementary sources are
 GRBs. The total UHECR luminosity of a galaxy due to GRBs is
\begin{equation}
L^{\rm gal}_{\rm GRB}  \left(E > 5 \times 10^{18}~\eV \right) 
\sim \Gamma_{\rm GRB}^{\rm gal} ~ \left(\frac{E_{\rm GRB}}{10^{52}~
{\rm ergs} } \right ) \simeq 10^{36} \  {\rm erg} {\rm s}^{-1}~.
\end{equation}
This luminosity almost coincide  with that previously estimated for UHECRs emitted by galaxies
so that, even in this case, we do not expect an observable secondary emission from the ICM.  

\section{Insertion of synthetic turbulence at small length scale}

As we discussed in the above, the maximal resolution of the MSPH simulation is about 
$2\pi/k_{min} \sim 10~\kpc$.
This means that the simulation does not account for magnetic fluctuations possibly present
below this scale.  Beside to the turbulent cascade,  MHD turbulence at small scales may be due to  
a  number of  effects, e.g. galactic winds and galaxy motion. 
Since in the central region of GCs the Larmor radius of protons  with energy below $10^{19}~\eV$ is 
smaller than the spatial resolution of our simulation, the presence of strong
magnetic fluctuations below this scale might spoil  the validity of our previous results at low energies. 

In order to clarify this issue, we redetermined the delay time in the MSPH magnetic structure 
of a massive cluster to which we added  turbulent MFs  at small scale.
The extra component of the MF is generated by summing over a large number of 
randomly distributed plane waves with spherically symmetric direction and with random polarizations and 
phases. We follow here the approach presented in \cite{Jokipi}  where it was showed 
that in the limit of an infinite number of wave modes, the
turbulence is isotropic and spatially homogeneous.  The modelled turbulent component is given by
\begin{equation}
\delta {\bf B}(x,y,z)=\sum_{n=1}^{N_m}A(k_n)\ {\bf\hat{\xi}_n}\
exp(ik_n z_n'+i\beta_n) \label{waves}
\end{equation}
where $A(k_n)$ represents the amplitude of the
wave mode with wave number $k_n$, polarization ${\hat \xi}_n$, and
phase $\beta_n$.  Eq. (\ref{waves}) satisfies $\nabla {\bf B}(x,y,z)=0$. 
The maximal value of $k_n$ that we used is $2\pi/0.5~\kpc^{-1}$.
For the sake of simplicity we assume the extra component to have a  Kolmogorov spectrum.
Since the MSPH simulation predict a stepper spectrum, we think that this assumption may 
only lead to overestimate  the turbulent power at small scales hence the correction
to the delay time.

\begin{figure}[h!]
\centerline{\epsfxsize=10cm \epsfbox{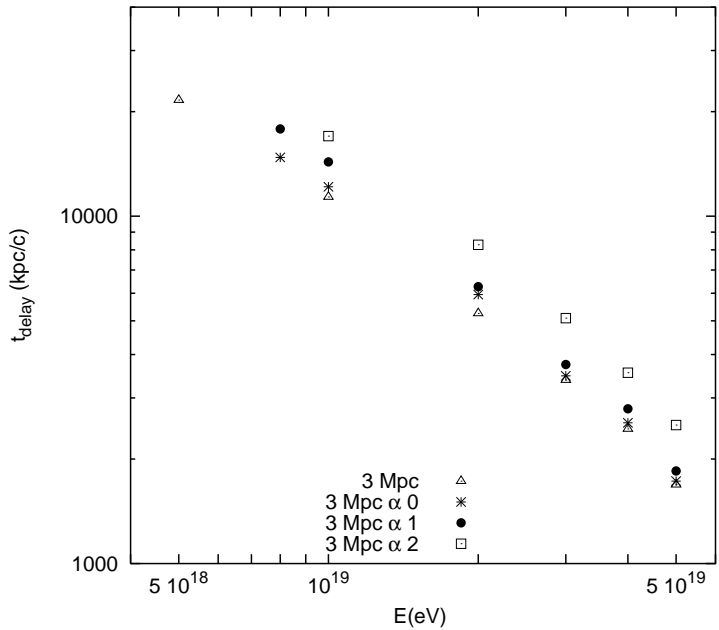}}
\caption{The same as Fig.\ref{tdelay} with an extra MHD turbulent power 
for several values of the parameters $r_0$ and $\alpha$ defined in Eq.\ref{extraturb}.}
\label{turb}
\end{figure}

For generality, we assume the extra  turbulent power to have the following radial dependence
\begin{equation}
\frac{{\bf \delta B}^2(r)}
{ {\bf B}_{\rm MSPH}^2(r_o)}=\left(\frac{r}{r_o}\right)^{-\alpha}~.
\label{extraturb}
\end{equation}
Indeed,  it is reasonable to assume 
that the level of turbulence will be higher in the center of 
GCs than outside. For example, since the formation of  GCs the
galaxies at the center make several revolutions around and may then
contribute to the level of MHD  turbulence.  
In this case, a resonable choice for $r_0$ is the cluster core radius $r_0 \sim 500~\kpc$.
We adopt this choice to compute the delay time of protons versus their energy.
This is plotted in Fig.\ref{turb} for different values of the parameter $\alpha$.
It is evident from this figure that the addition of turbulence at small scales
doesn't increase significantly the diffusion time. This means that diffusion properties  of protons
in the energy range $10^{18} - 10^{20}~\eV$ are dominated by the MF fluctuations 
at larger scale, where most of the power is concentrated. Since these fluctuations  are well 
accounted by the MSPH simulation we can reasonably trust the results presented in Sec. 3.

\section{Diffusion of UHE protons, comparison with previous results}

Diffusion of cosmic rays in disorganised magnetic fields takes place by scattering of the
charged particles onto the magnetic irregularities. Particles interact with
the field when their gyro-motion resonates with a Fourier component of 
wavelength equal to the Larmor radius.
The diffusion coefficient, defined by $\displaystyle D(E) \equiv \frac{\langle \Delta x^2 \rangle}
{ 2 \Delta t}$, can be computed  analytically in the regime of weak turbulence and for
low values of the magnetic rigidity $\rho \equiv 2\pi r_L /L_c \ll 1$. 
In this case 
\begin{equation}
D(E) \simeq \frac{1}{3} r_L c \frac {\langle B^2 \rangle }{\int_{1/r_L}^{\infty}
dk B^2(k)k^2 } \ ,
\label{diffcoef}
\end{equation}
where $B^2(k)k^2$ is the magnetic field energy power spectrum. 
In the case of a Kolmogorov spectrum, $B^2(k)k^2 \propto k^{-5/3}$ , the energy dependence 
$D(E) \propto E^{-1/3}$ is obtained.   A priori there are no reasons why   Eq.(\ref{diffcoef}) 
should hold for UHE protons diffusing in the ICMFs. In  Sec. 2 we showed
that MSPH predict strong turbulence ($\eta = 0.6$)  in the core of rich clusters
($R < 500~\kpc$)  with a power spectrum which may be significantly steeper than a Kolmogorov's. 
Furthermore, since the Larmor radius  
\begin{equation}
\ r_L \simeq  \frac{E}{ZeB} \simeq  10  
\left(\frac{E}{10^{19}~\eV}\right) \left(\frac{B}{10^{-6}~{\rm G}}\right)^{-1}\, \kpc ,
\label{Larmor}
\end{equation}
is comparable to  the field coherence length $L_c$, the magnetic rigidity of UHE protons in the cluster is quite close to unity. 

Numerical simulations of charged particle diffusion have been performed in the case of strong Kolmogorov turbulence  \cite{Pelletier}. These simulations confirmed the validity of Eq.(\ref{diffcoef}) for low values of the magnetic rigidity.  However, significant
 deviations from the $ E^{-1/3}$ behaviour were found  when the magnetic rigidity approaches unity. The best fit to the numerical results gives 
(in units of $\Mpc^2/{\rm Myr}$):
 \begin{eqnarray}
 D(E) &\simeq& 0.004\left(\frac{E}{10^{20}~\eV} \right)^{1/3}
\left(\frac{B}{\muG} \right)^{-1/3}\left(\frac{L}{\Mpc}\right)^{-2/3},  
\quad E \simleq 0.1E_{\rm cr} \nonumber\\
D(E) &\simeq& 0.03\left(\frac{E}{10^{20}~\eV} \right)\left(\frac{B}{\mu G} 
\right), \qquad 0.1 E_{cr} \simleq E \simleq E_{\rm cr}  
\label{pelleq}\\
D(E) &\simeq& 0.02\left(\frac{E}{10^{20}~\eV} \right)^{7/3}\left(\frac{B}{\mu G}
 \right)\left(\frac{L}{Mpc}\right)^{-4/3},  \quad E_{\rm cr} \simleq E \nonumber
\end{eqnarray}
In these expressions $E = E_{\rm cr}$ corresponds to the condition $\rho =1$. 
Accordingly  to these results,  the energy dependence of the diffusion time, $t_{\rm diff}(E) = r^2/6 D(E)$,   changes rather smoothly from a quasi-rectilinear regime at $E \simeq 
E_{\rm cr}$, with $t_{\rm diff}(E) \propto E^{-7/3}$,  to the, so called, Bohm diffusion 
regime with $ t_{\rm diff}(E) \propto E^{-1}$ , for $0.1 E_{cr} \simleq E \simleq E_{\rm cr}$.
The analytically predicted behaviour, $t_{\rm diff}(E) \propto E^{-1/3}$,  was found 
only at  low energies, $E \simleq 0.1E_{cr}$.

However, the properties of the magnetic field configuration assumed in \cite{Pelletier} are 
still quite different from those of ICMF given  by our MSPH simulation. 
 Indeed,  the mean field intensity, which was assumed to be uniform in \cite{Pelletier},  
has a quite pronounced radial profile in all the simulated clusters.  Furthermore, MSPH simulations 
predict a turbulent power spectrum which is often quite different from  
a Kolmogorov's.  We think that these simulation provide a much more realistic picture
of the real ICMF configuration. 

First of all we checked that our numerical code reproduces the diffusion coefficients 
found in \cite{Pelletier} in the energy range $10^{18}-10^{20}~\eV$ for strong homogeneous 
turbulence.  Remarkably, we found that our results are in good agreement with those of \cite{Pelletier} 
also in the case the field configuration is that determined by the MSPH simulation.
 The energy at which the delay time passes from a $\sim E^{-2}$ dependence, consistently  with
 small pitch angle diffusion, to $~E^{-1}$ (Bohm scattering diffusion) is  $\sim 3 \times 10^{19}~
 \eV$.  This energy corresponds to a Larmor radius of $\sim 30\kpc$, 
for $\langle B \rangle  \simeq  1~\muG$, which almost coincide with the scale 
 $L_{min} = 2\pi /k_{min}$ at which we found that most of ICMF energy is concentrated.

\section{Hard X and gamma-ray emissions}

In this section we apply our previous results  to estimate the flux of
high energy photons  produced by the interaction of  UHE protons  with the gas and the 
radiation in the magnetised core of a  rich GCs. Such a radiation may provide a signature
of the possible presence of  powerful UHECR sources in galaxy clusters. 
For the sake of simplicity we will assume that a GC harbour a single point-like 
 UHE proton source placed at its center.  We will only consider the contribution of protons in the 
energy range  $5 \times 10^{18} \simleq E \simleq 3 \times 10^{19}~\eV$ with 
injection spectrum  $n_p(E) dE = E^{-\gamma} dE$
\footnote{By injection spectrum we  mean here the spectrum of protons which
leave the acceleration region reaching the ICM. This may differ from the 
acceleration spectrum due to energy losses close to  the source.}.

We start  estimating  the flux of  gamma-rays due to hard hadronic interactions ($pp$-scattering) of UHE protons with the baryon gas in the ICM. 
The differential photon production rate $Q_\gamma(E)$  
from the decay of secondary $\pi^0$'s  is given by  \cite{Bere97}:
\begin{equation}
Q_{\gamma}(E) \simeq Y_{\gamma}~\frac{L_p(E > E_{\rm min})}{E_{\rm min}^2}
\left( \frac{E}{E_{\rm min}} \right)^{-\gamma} \sigma_{pp} n_{\rm gas} c t_{\rm diff}(E)
\end{equation}
where $Y_{\gamma}$ is the fraction of proton energy transferred to the photons
 and $\sigma_{pp} \simeq 3 \times 10^{-26}~{\rm cm}^2$ is the 
$pp$-scattering cross section (we neglect here a weak energy dependence of this quantity). 
In Sec.3 we found that, 
in the energy range $5 \times 10^{18} \simleq E \simleq 5 \times 10^{19}~\eV$,
the diffusion path length of protons in the core of a cluster with mass  
$\sim 8 \times 10^{14}~{\rm M}_\odot$ (like Coma) is
\begin{equation}
\label{ctdiff}
 c t_{\rm diff}(E) \simeq  20
 \left( \frac {E}{5 \times 10^{18}~\eV} \right)^{-1} \  \Mpc ~.
\end{equation}
Therefore the gamma-ray energy emission rate per logarithmic energy interval of the primary 
protons is 
\begin{equation}
\label{E2Q}
E^2  Q_{\gamma}(E) \simeq 10^{40} 
\left(\frac{Y_\gamma}{0.1}\right)
\left( \frac{n_{\rm gas}}{10^{-3}~{\rm cm}^{-3}} \right) 
\left( \frac{L_p \left( E > E_{\rm min}\right)} {10^{44}~{\rm erg~s}^{- 1}} \right)
 \left(\frac{E}{E_{\rm min}}\right)^{1-\gamma}{\rm erg~s}^{-1} , 
\end{equation}
where $E_{\rm min} = 5 \times 10^{18}~\eV$.
 Since the mean free path of $E_\gamma \sim 10^{19}~\eV$ photons,  due to 
 $\gamma + \gamma_{\rm CMB} \rightarrow e^++e^-$,  is much smaller than the  cluster
 core radius \cite{Prothe},  practically  the entire  energy of the photons  will be converted into
electromagnetic showers inside the cluster. 
The shower, however,  cannot escape the cluster.  This is promptly understood by comparing
the synchrotron energy loss length of  the  electron-positron component of the shower,
\begin{equation}
l_{\rm syn} \simeq 2 \left(\frac{B}{1~\muG} \right)^2
\left(\frac{10^{18}eV}{E_e}\right) ~ {\rm pc}~, 
\nonumber
\end{equation}
with the cluster size and  the  IC scattering length over which electrons and positrons might energise  
CMB photons, which are much larger.
 Therefore, all the energy  of the  photons produced by $\pi^0$ decay is converted into synchrotron 
photons. For monochromatic electrons, and for a uniform magnetic field, the photon synchrotron spectrum 
would peak at the energy   
\begin{equation}
\hbar \omega_{\rm syn} = \frac{3\sqrt{\alpha}B}{2m_e^3}E^2  \simeq
10  \left(\frac{E}{10^{18}eV} \right)^2\left(\frac{B}{1~\muG}\right)   \GeV
\label{Esyn}
\end{equation} 
with  width $\delta \omega \sim \omega_{\rm syn}$. 
Secondary electrons (and positrons) from $pp$-scattering,  however, are spread  over a range
of frequencies which is much larger than $\delta \omega$ so that the spectral shape of the 
synchrotron emission is mainly determined (we neglect here a possible dependence on the IGMF power 
spectrum which will be investigate elsewhere) by the electron energy spectrum\cite{Longair} hence, 
in turn,  by the primary proton spectrum in the cluster.  

It is worth noticing here,  that the synchrotron emission in the GeV region would be overwhelmed by the secondary emission produced 
by low energy protons  if the UHE protons spectrum extends to lower energies with a power  $\gamma \simgeq 2$ .
There may be cases, however, in which the proton emission is peaked at ultra high energies
 (see below about possible scenarios which may give rise to this situation). 

If we assume that the proton energy spectrum in the ICM peaks at $E_{\rm max}\sim 10^{19}~\eV$
 the synchrotron emission will peak at an energy of few tens of GeV's.
Photons with this energy can travel over cosmological distances without undergoing significant
 energy losses and might be detectable by the GLAST   satellite \cite{GLAST} or the MAGIC 
 Cherenkov telescope \cite{MAGIC}.   We find, however, that the expected photon flux 
\begin{equation}
{\dot n}_\gamma  \simeq  \frac{1}{E_\gamma}~\frac{E^2 Q_\gamma(E)}{4\pi d^2} \simeq
3 \times 10^{-13}  \left(\frac{L_p\left( E >  5 \times 10^{18}~\eV \right)}
{10^{44}~{\rm erg~s}^{-1}} \right) \left( \frac{d}{100~\Mpc}\right)^2
 {\rm cm}^2  s^{-1}~
\end{equation}
is much lower than the expected sensitivity of these instruments, unless the UHE proton source 
is extremely bright  $\left( L_p \left( E >  5 \times 10^{18}~\eV \right) \gg  10^{46}~{\rm erg~s}^{-1} \right) $ which, however, may contradict other constraints.

A more promising  signature of the presence of a powerful UHE proton source in a GC  could be given by the synchrotron emission of secondary electrons and positrons produced by the process $p + \gamma_{CMB} \rightarrow e^++e^-+p$ (proton photo-pair production).
It is remarkable that,  for protons in the energy range $5 \times 10^{18} \simleq E \simleq 3 \times 10^{19}~\eV$, the secondary electrons produced by this process are practically 
monochromatic with  energy $\sim 830$ TeV \cite{Berezinsky}. As a consequence,  the spectrum of the synchrotron photons emitted in the cluster peaks in the hard X-ray (HXR) range,  around 
$\sim 10 ~\keV$  (see Eq.\ref{Esyn}). 
Interestingly,  a  non-thermal emission from the Coma cluster, peaking  at this energy,
 has been  already detected. 
The observed energy flux  is  \cite{Fusco,Fusco03} 
$E_X^2 f(E_X) \simeq 10^{-11}~{\rm erg}~{\rm cm}^2  s^{-1}$.
In order to estimate the primary proton flux required to explain this signal in 
terms of secondary $e^\pm$ synchrotron emission we need to know  the probability
 for  a proton to undergo photo-pair production scattering inside the cluster.  
This probability is maximal at $E \sim 10^{19}~\eV$, at which it takes the value  
\begin{equation}
\label{Ppair}
{\rm P}_{\rm pair}(10^{19}~\eV)  = 
 1 -  \exp \left( - \frac{ c t_{\rm diff}(10^{19}~\eV) }
{ l_{\rm pair}(10^{19}~\eV) } \right) \simeq  10^{-2} \ ,
\end{equation} 
where we used our Eq. (\ref{ctdiff}) to find the proton diffusion time in the 
cluster and the interaction length 
$ l_{\rm pair}(E \sim 10^{19}~\eV) \simeq 1000~ \Mpc$ 
determined in \cite{Blume,Chodo}.
Therefore the UHE proton luminosity required to explain the Coma HXR emission 
entirely in terms of synchrotron emission of electrons produced by proton pair production 
scattering is given by
\begin{equation}
\label{Lpx}
L_p  \left( E > 5 \times 10^{18}~\eV \right)  
= 4\pi d^2 ~{\rm P}_{\rm pair}^{-1}~ \Phi_X \simeq  10^{45}~\ergs~.
\end{equation}
This luminosity corresponds to $L_p  \left( E > 1~\GeV \right)  
\simgeq  10^{46}~\ergs$ for $\gamma \simgeq 2$, and it increases rapidly  for larger values of $\gamma$.
Even for $\gamma \simeq 2$,  the required proton luminosity would be too high to be
compatible with the EGRET limit \cite{EGRET} on the secondary gamma-ray emission  in the  
100 Mev - 10 GeV range due to  $pp$-scattering .  It would also give rise to a too intense radio synchrotron emission from $pp$ secondary electrons \cite{Blasi99}. 
 Furthermore,   a proton source with  luminosity 
$L_p  \left( E > 5 \times 10^{18}~\eV \right)  > 10^{45}~~\ergs$ may be at
odd with the results of UHECR experiments which, so far, do not show any evidence of a
flux excess in the direction of Coma. 
 
A possible way out from some of these problems is to assume a flatter proton spectra in the cluster,  
i.e. $\gamma < 2$.
  This may be the case if UHE protons are accelerated in the knots and hot spots of  a  powerful AGN's
 jet \cite{Aharonian01}.   
 Another possibility is that the proton acceleration  is not stochastic. Rather, it may be induced  by 
the huge electric field generated by a supermassive rotating black-hole (a so called ``dead quasar").
It was showed in \cite{Neronov04} that proton energies as large as $5 \times 10^{19}~\eV$ can be reached 
 in this case with a quite narrow spectral distribution.  Although, a quite intense TeV gamma-ray
direct emission  is expected from such an object,  this emission is predicted to be beamed so that 
its detection may be missed.  In a such a case the  X and gamma-ray diffuse emission 
produced by secondary particles in a  GC harbouring the source may offer an independent 
signature.   

Even if  the HXR excess from Coma is not due to  an intense UHE protons emission in that GC,  weaker 
proton sources may still give rise to a detectable HXR signal from closer GCs. Indeed the expected 
X-ray flux in the $10-100~\keV$ region due to synchrotron
emission of secondary electrons is 
\begin{equation}
\Phi_X \simeq 2 \times 10^{-12} ~\left( \frac{{\rm P}_{\rm pair}}{10^{-2}}\right)
\left( \frac{d}{20~\Mpc}\right)^2~
\left(\frac{ L_p  \left( E > 5 \times 10^{18}~\eV \right) }{10^{43}~\ergs}\right)
~~\ergscm2
\end{equation}
A proton luminosity $L_p  \left( E > 5 \times 10^{18}~\eV \right)  \simeq 10^{43}~\ergs$
corresponds to $L_p  \left( E > 1~\GeV \right)  \simeq 10^{44}~\ergs$ for $\gamma = 
2.1$.   This is a standard luminosity for an AGN. The required luminosity
can be even smaller if the proton spectrum is peaked at high energies. 
Therefore an AGN or a dead quasar accelerating protons at UHEs  may give rise to a detectable 
signature under the form of synchrotron HXRs if they are harboured by a rich GC in the 
local supercluster. 
 
We note on passing, that a  mechanism similar to that we just discussed  here has been recently
considered in \cite{Neronov03}.   The main difference with our  mechanism is that in 
\cite{Neronov03}  the synchrotron emitting electron-positron pairs were assumed to be
 produced by photons with energy $E_\gamma = 700$ TeV.  No  compelling process was 
 suggested, however,  to explain the origin of the required flux of photons with that particular
 energy.

An independent signature of the presence of a UHE proton bright source  may be provided by the 
secondary gamma ray
radiation produced  in the IGM by the protons  which escape the clusters.
The relevant process at $E \simgeq 10^{19}~\eV$ is, again, proton pair-production onto the CMB.   
If IGMFs are weaker than $10^{-10}$ G, as it is suggested by the results
of the large scale MSPH simulation performed in \cite{Dolag03},    the electromagnetic
showers produced by the secondary electrons and photons can travel over  a distance of  hundreds Mpc's
\cite{Ferrigno04}.  For a source at a distance of $\sim 20~\Mpc$  the electromagnetic shower 
at the observer position will be composed by photons, electrons and positrons with energies 
which can exceed  several  TeV's.
Since the shower  production probability is in this case  $P_{\rm pair} \sim 2 \times 10^{-2}$,  
the photon flux around 1 TeV can be roughly estimated to be
\begin{equation}
{\dot n}_\gamma(E_\gamma\sim 1~\TeV) = \frac{L_p}{ 4\pi d^2~ E_\gamma}  
{\rm P}_{\rm pair}  \simeq  2 \times 10^{-12}   
 \left(\frac{L_p}{10^{43}~{\rm erg~s}^{-1}} \right) 
\left( \frac{d}{20~\Mpc}\right)^2 {\rm cm}^2 {\rm s}^{-1}~.
\end{equation}
Such a flux should not be missed by Cherenkov gamma-ray telescopes like 
MAGIC \cite{MAGIC}, HESS \cite{HESS} and VERITAS \cite{VERITAS}.  

Detailed computations of the expected spectra in the gamma as well as in the HXR region 
are beyond the aims of this work and they will  be presented elsewhere.

\section{Conclusions}

In this paper we investigated the propagation of UHE protons in the magnetised  medium
of Galaxy Clusters.   By using a constrained  MSPH simulation of the magnetic field structure
in the local universe we were able to account, for the first time,  for several features of the ICMF  
which we think to be  present in actual nearby clusters.   
We showed that UHE propagation takes place in the spatial diffusion regime in the core of rich GCs
($M \simeq 10^{15}~M_\odot$)  up to the $E_{\rm cr} \simeq  3 \times 10^{19}~\eV$.
Below this energy, at least down to $\sim 5 \times 10^{18}~\eV$, the diffusion time scales with energy
like $E^{-1}$ (Bohm scattering diffusion).  
In spite of the different  ICMF power spectrum found in different simulated clusters, 
we found that the behaviour of the proton residence time as a function of  the energy does
not changes significantly from cluster to cluster. Only the amount of the delay  changes 
depending on the mean intensity of the ICMF. 
The path length increase due to the ICMF is too small to give rise to a significant modification of 
the proton energy spectrum due to energy losses in the ICM, since $c t_{\rm delay} \sim 10~\Mpc \ll
l_{\rm loss} \sim 10^3~\Mpc$ for $E \sim 10^{19}~\eV$.
Since the residence time of UHE protons is much larger than that expected for  straight propagation, 
and for small pitch angle diffusion, the probability for them to undergo hadronic or photo-pair 
production scattering is considerably increased. 
 In order to determine a possible signature of a bright proton source harboured by a GC, 
 we applied our results to estimate  the gamma ray and HXR secondary emission produced by UHE protons
in the ICM. We showed that electromagnetic showers produced by $pp$-scattering and proton photo-pair 
production do not leave the cluster due to the intense synchrotron losses. Therefore, their energy is
transferred to synchrotron photons of lower energy which can reach the observer without further losses.
The synchrotron gamma-ray emission from the electrons produced by the decay of secondary pions is too 
weak to be detected.
 More promising is the synchrotron emission of secondary electrons and 
positrons produced by proton photo-pair scattering which falls in the HXR range.  We showed that
 the UHE proton emission of a relatively powerful AGN placed in a GC in the local supercluster
may give rise to a detectable HXR  emission.

\section*{Acknowledgments}

We would like to thank P. Blasi and M. Vietri for helpful discussions and suggestions.
We also thank V. Springel and I. Tkachev for collaboration performing the constrained MSPH 
simulation which we used in this paper. 
D.G. thanks Lin Yin for collaboration in the initial phase of this work. 
K. Dolag acknowledges support by a Marie Curie Fellowship of the European Community program 
'Human Potential'  under contract number MCFI-2001-01227.


\begin{thebibliography}{9}

\bibitem{Berezinsky}
V. ~S.~Berezinsky {\it et al.},
`` Astrophysics of Cosmic Rays", North-Holland, Amsterdam, 1990.


\bibitem{Bere97} V.~S.~Berezinsky, P.~Blasi and V.~S.~Ptuskin,
Astroph. \ J. \ {\bf 487}, 529 (1997)
[arXiv:astro-ph/9609048].


\bibitem{Feretti99} 
L.~Feretti, in ``Diffuse thermal and relativistic plasma in galaxy clusters", 
eds. Bohringer et al. MPE report {\bf 271}, 3 (1999).

\bibitem{Giovannini93}
G.~Giovannini {\it et al.}, 
Astroph. \ J. \  {\bf 406}, 399 (1993).

\bibitem{Fusco}
R.~Fusco-Femiano {\it et al.},
Astroph. \ J.\  {\bf 513}, L21 (1999).
[arXiv:astro-ph/9901018].

\bibitem{Rephaeli99}
Y.~Rephaeli, J.M.~Stone, and P.~Blanco,
Astroph. \ J.\  {\bf 511}, 21 (1999).
 
\bibitem{Blasi99}
P.~Blasi and S.~Colafrancesco,
Astropart.\ Phys.\  {\bf 122} (1999) 169
[arXiv:astro-ph/9905122].

 

\bibitem{Atoyan99}
A.~M.~Atoyan and H.~J.~Volk,
Astrophys.\ J.\ {\bf 535}, 45 (2000)
[arXiv:astro-ph/9912557].

\bibitem{Neronov03}
A.~N.~Timokhin, F.~A.~Aharonian and A.~Y.~Neronov,
Astron.\ Astrophys.\  {\bf 417}, 391 (2004)
[arXiv:astro-ph/0305149].

\bibitem{Feretti99b}
L.~Feretti {\it et al.},
Astron.\ Astrophys.\  {\bf 344}, 472 (1999)
[arXiv:astro-ph/9902019].

\bibitem{Taylor01}
G.~B.~Taylor, {\it et al.},
Mon.\ Not.\ Roy.\ Astron.\ Soc.\  {\bf 326} (2001) 2
[arXiv:astro-ph/0104223].

\bibitem{Carilli02}
C.~L.~Carilli and G.~B.~Taylor,
Ann.\ Rev.\ Astron.\ Astrophys.\  {\bf 40} (2002) 319
[arXiv:astro-ph/0110655].

\bibitem{Fusco03}
R.~Fusco-Femiano, {\it et al.}, 
Astrophys.\ J.\  {\bf 602} (2004) L73
[arXiv:astro-ph/0312625].

\bibitem{Vogt03}
C.~Vogt and T.~A.~Ensslin,
arXiv:astro-ph/0309441.

\bibitem{Murgia04}
M.~Murgia {\it et al.}, submitted to Astron.\ \& Astrophys., 2004.

\bibitem{Dolag03} 
K.~Dolag, D.~Grasso, V.~Springel and I.~Tkachev,
arXiv:astro-ph/0310902.

\bibitem{Dolag02}  
K.~Dolag, M.~Bartelmann and H.~Lesch,
Astron.\ \& Astrophys. {\bf 387}, 383 (2002).
arXiv:astro-ph/0202272.

\bibitem{report}
D.~Grasso and H.~R.~Rubinstein,
Phys.\ Rept.\  {\bf 348} (2001) 163
[arXiv:astro-ph/0009061].

\bibitem{Furlanetto}
S.~Furlanetto and A.~Loeb,
Astrophys.\ J.\ {\bf 556}, 619 (2001)
[arXiv:astro-ph/0102076].

\bibitem{Volk98}
H.~J.~Volk and A.~M.~Atoyan,
Astrophys.\ J.\ {\bf 541}, 88 (2000)
[arXiv:astro-ph/9812458].

\bibitem{battery}
D.~Ryu, H.~Kang and P.~L.~Biermann,
Astron.\ \& Astrophys. {\bf 335}, 19 (1998).

\bibitem{Kulsrud97}
R.~M.~Kulsrud, R.~Cen, J.~P.~Ostriker and D.~Ryu,
Astrophys.\ J.\ {\bf 480}, 481 (1997)
[arXiv:astro-ph/9607141].

\bibitem{Dolag99}
K.~Dolag, M.~Bartelmann and H.~Lesch,
arXiv:astro-ph/9906329.

\bibitem{Mathis} H.~Mathis et al.,
MNRAS, {\bf 333}, 739 (2002).

\bibitem{Waxman}
E.~Waxman,
Phys.\ Rev.\ Lett.\  {\bf 75} (1995) 386
[arXiv:astro-ph/9505082].

\bibitem{Vietri}
M.~Vietri,
Astrophys.\ J.\  {\bf 453} (1995) 883
[arXiv:astro-ph/9506081].

\bibitem{Jokipi}
J~.Giacalone and J.R.~Jokipii
Astrophys.\ J.\ {\bf 520}, 204 (1999). 

\bibitem{Pelletier}
F.~Casse, M.~Lemoine and G.~Pelletier,
Phys.\ Rev.\ D {\bf 65} (2002) 023002
[arXiv:astro-ph/0109223].

\bibitem{Prothe}
R.~J.~Protheroe and P.~L.~Biermann,
Astropart.\ Phys.\  {\bf 6} (1996) 45
[Erratum-ibid.\  {\bf 7} (1997) 181]
[arXiv:astro-ph/9605119].

\bibitem{Longair}
M.~S.~Longair, ``High Energy Astrophysics", Cambrigde Press 1994.




\bibitem{GLAST} GLAST: http://glast.gsfc.nasa.gov/

\bibitem{MAGIC} MAGIC: http://hegra1.mppmu.mpg.de/MAGICWeb/


\bibitem{EGRET} P.~Sreekumar {\it et al.},
Astrophys.\ J.\ {\bf 464}, 628 (1996).

\bibitem{Aharonian01} 
F.~A.~Aharonian,
Mon.\ Not.\ Roy.\ Astron.\ Soc.\  {\bf 332} (2002) 215
[arXiv:astro-ph/0106037].

\bibitem{Neronov04}
A.~Neronov, P.~Tinyakov and I.~Tkachev,
arXiv:astro-ph/0402132.



\bibitem{Blume}
G.~R.~Blumenthal,
Phys.\ Rev.\ D {\bf 1} (1970) 1596.

\bibitem{Chodo}
M.J.~Chodorowski, A.A.~Zdziarsky and M.~Sikora,
Astrophys.\ J.\ {\bf 400}, 181 (1992). 

\bibitem{Ferrigno04}
C.~Ferrigno, P.~Blasi and D.~De Marco,
arXiv:astro-ph/0404352.

\bibitem{HESS} HESS: http://www.mpi-hd.mpg.de/hfm/HESS/HESS.html

\bibitem{VERITAS} VERITAS: http://veritas.sao.arizona.edu/


\end{thebibliography}
\end{document}